\documentclass{ws-ijmpa}

\usepackage[super]{cite}
\usepackage{xcolor}
\usepackage[verbose,hypertexnames=false]{hyperref}
\hypersetup{colorlinks=false,allbordercolors=blue,pdfborderstyle={/S/U/W 1}}

\begin{document}

\markboth{Authors' Names}{Instructions for typing manuscripts (paper's title)}

%
\catchline{}{}{}{}{}
%

\title{Particle Physics in Curved Spacetime and Dark Matter}

\author{Antonio Capolupo, Gabriele Pisacane}

\address{Dipartimento di Fisica ``E.R. Caianiello'', Universit\`{a} di Salerno and INFN -- Gruppo Collegato di Salerno, Via Giovanni Paolo II, 132\\
	Fisciano, SA 84084, Italy\\
	capolupo@sa.infn.it, gpisacane@unisa.it}

\author{Salvatore Capozziello}

\address{Dipartimento di Fisica "E. Pancini", Universit\`a degli Studi di Napoli Federico II", Via Cinthia, I-80126, Napoli, Italy and Scuola Superiore Meridionale, Largo S. Marcellino 10, I-80138, Napoli, Italy.\\
capozziello@na.infn.it}

\author{Aniello Quaranta}

\address{School of Science and Technology, University of Camerino, Via Madonna delle Carceri\\
	Camerino, MC 62032, Italy\\
	aniello.quaranta@unicam.it}

\maketitle

\begin{history}
\received{Day Month Year}
\revised{Day Month Year}
\accepted{Day Month Year}
\published{Day Month Year}
\end{history}

\begin{abstract}
We review recent results showing that, within the framework of quantum field theory in curved spacetime, the semiclassical energy-momentum tensor of the neutrino flavor vacuum fulfills the equation of state of dust and cold dark matter. By considering spherically symmetric spacetimes in the weak field approximation, the flavor vacuum is shown to contribute as a Yukawa correction to the Newtonian potential. We discuss how this modified potential provides a mechanism to account for the flat rotation curves of spiral galaxies. In this perspective, neutrino mixing is presented as a viable contributing factor to the dark matter content of the universe \cite{principal}.
\end{abstract}


\ccode{PACS numbers: 03.65.$-$w, 04.62.+v}

\section{Introduction}

In the framework of the $\mathrm{\Lambda CDM}$ model, modern cosmology accounts for the observed universe by introducing dark energy to explain the accelerated expansion \cite{DE1}, and non-baryonic dark matter to justify the flat rotation curves of spiral galaxies and the gravitational stability of large scale structures \cite{Dm1, Dm2, Dm3, Dm4, Dm5, Dm6, Dm7, Dm8, Dm9, Dm10, Dm11, Dm12, Dm13, Dm14, Dm15, Dm16, Dm17, Dm18}. Although these dark components constitute the majority of the total energy density of the universe, their composition and origin remain open issues. Several beyond the standard model theories have been proposed to explain dark matter through hypothetical particles, such as axions \cite{Axion1, Axion2, Axion3, Axion4, Axion5, Axion6, Axion7, Axion8, Axion9, Axion10, Axion11}, mirror matter \cite{MM1, MM2, MM4, MM7, R5}, and sterile neutrinos \cite{SN}. Alternative interpretations rely on extended theories of gravity \cite{ETG1} or modifications to General Relativity \cite{Cai}. Furthermore, the interplay between background geometry and QFT vacuum effects, including gravitational particle production and the subsequent generation of quantum entanglement in cosmological backgrounds, has been recently investigated to provide novel perspectives on the dark sector and cosmological dynamics \cite{Belfiglio1, Belfiglio2, Belfiglio3, Belfiglio4, Belfiglio5, Belfiglio6, Belfiglio7}. On the other hand, it is a well-established fact that massive neutrinos \cite{N1, N5} play a significant role in astrophysical and cosmological contexts \cite{CN1}. Representing a direct challenge to the standard model of particle physics, neutrino mixing carries numerous phenomenological implications. The quantum field theory (QFT) of neutrino mixing involves a non-trivial vacuum state, known as the flavor vacuum \cite{N8, FV1}. In virtue of its condensate structure, this vacuum can act as an additional gravitating source, emerging directly from QFT \cite{Capolupo2016}. In this work, we show the properties of the energy-momentum tensor associated with the flavor vacuum. Generalizing previous results obtained in flat space \cite{Capolupo2016} and in Friedmann-Lemaitre-Robertson-Walker (FLRW) metrics \cite{FV2}, it is discussed how this tensor acquires a perfect fluid form with an equation of state typical of cold dark matter. By analyzing the flavor vacuum in static spherically symmetric metrics, we report the derivation of the induced gravitational potential in the weak field approximation. Finally, we discuss how this Yukawa-like correction can fit the dynamics of spiral galaxies and the baryonic Tully-Fisher relation, suggesting that neutrino mixing may generate corrections to the gravitational potential acting at galactic scales \cite{Milgrom}.
\section{Particle Mixing in QFT and the Flavor Vacuum}
Let us review the quantization of flavor fields in curved spacetime \cite{FV1, FV2}. Considering a two-flavor neutrino mixing framework, the action for the free Dirac fields $\nu_1 (x)$ and $\nu_2 (x)$ with masses $M_1$ and $M_2$ is given by:
\begin{equation}\label{Eq.:Action}
	S = \sum_{L=1,2} \int d^4 x \ \sqrt{-g} \left[ \frac{i}{2} \left(\bar{\nu}_L \tilde{\gamma}^{\mu} D_{\mu} \nu_L - D_{\mu} \bar{\nu}_L \tilde{\gamma}^{\mu} \nu_L \right)-M_L \bar{\nu}_L \nu_L \right] \ .
\end{equation}
where $\tilde{\gamma}^{\mu} = e^\mu_A \gamma^A$ are the curved space Dirac matrices, defined via the tetrad fields $e^\mu_A$, and $D{\mu}$ denotes the spinorial covariant derivative. Expanding these mass fields in terms of a complete, orthonormal set of positive and negative energy solutions to the Dirac equations establishes the mass vacuum $|0\rangle$.The physical flavor fields, $\nu_e (x)$ and $\nu_\mu (x)$, are obtained through a rotation of the mass fields parameterized by the mixing angle $\Theta$:
\begin{equation}
	\begin{pmatrix} \nu_e (x) \\ \nu_\mu(x) \end{pmatrix} = \begin{pmatrix} \cos \Theta & \sin \Theta \\ - \sin \Theta & \cos \Theta \end{pmatrix} \begin{pmatrix} \nu_1 (x) \\ \nu_2(x) \end{pmatrix} \ .
\end{equation} 
In the quantum field theory formulation, this transformation is dynamically implemented by a time-dependent mixing generator $\mathcal{J}_{\Theta} (\tau)$. The action of this generator on the mass ladder operators yields a Bogoliubov transformation nested into a rotation \cite{FV1, FV2, FV3}. Consequently, the flavor annihilation operators define a distinct, time-dependent state known as the flavor vacuum:\begin{equation}|0_F (\tau)\rangle = \mathcal{J}^{-1}{\Theta} (\tau) |0\rangle \ .
\end{equation}
This state is unitarily inequivalent to the mass vacuum and possesses the structure of a condensate of particle-antiparticle pairs with definite masses. Due to this condensate structure, the energy-momentum content of the flavor vacuum is non-trivial. The classical energy-momentum tensor derived from the action \eqref{Eq.:Action} is:
\begin{equation}
	T{\mu \rho} (x) = \frac{i}{2}\sum_{L=1,2} \left[ \bar{\nu}L \tilde{\gamma}{(\mu} D_{\rho)} \nu_L - D_{(\mu} \bar{\nu}L \tilde{\gamma}{\rho)} \nu_L\right] \ .
\end{equation}The relevant physical quantity is its vacuum expectation value (VEV) evaluated on the flavor vacuum. To isolate the pure mixing contribution and remove the standard zero-point energy, the tensor is normal-ordered with respect to the mass vacuum:
\begin{equation}
	\mathbb{T}{\mu \rho} (x) = \langle 0_F (\tau) | : T{\mu \rho} (x) : | 0_F (\tau) \rangle \ .
\end{equation}
	A general algebraic property of $\mathbb{T}_{\mu \rho}$ is that any bilinear combination of the fields involving solely spatial derivatives is invariant under the action of the mixing generator $\mathcal{J}_{\Theta} (\tau)$. Consequently, all the spatial components of the energy-momentum tensor VEV vanish identically, yielding $\mathbb{T}_{jk} = 0$ for $j, k = 1,2,3$. Components involving time derivatives, such as $\mathbb{T}_{00}$ and $\mathbb{T}_{0i}$, are generally non-zero. However, it can be demonstrated that the off-diagonal components $\mathbb{T}_{0i}$ vanish in relevant symmetric spacetimes (such as FLRW or static spherically symmetric metrics), leaving only the energy density component $\mathbb{T}_{00} \equiv \rho$. Therefore, the energy-momentum tensor associated with the flavor vacuum attains the perfect fluid form with vanishing pressure ($p=0$):
	\begin{equation}
		p = 0 = w \rho \quad \implies \quad w = 0 \ .
	\end{equation}
	This implies that the flavor vacuum is strictly characterized by the equation of state of dust and cold dark matter.To explicitly evaluate the energy density component $\mathbb{T}_{00}$ at galactic scales, one considers a static spherically symmetric background in isotropic coordinates:
	
	\begin{equation}
	ds^2 = f(R) dt^2 - g(R) \left(dx^2 + dy^2 + dz^2 \right) \ .
	\end{equation}
	Solving the Dirac equation in this metric requires expanding the spinor fields using an ansatz that separates the temporal, radial, and angular dependencies. The positive-energy solutions are characterized by two-component spherical spinors $\chi_{\kappa, m_j}$ and by the real radial functions $\Phi_{L}(R)$ and $\Psi_{L}(R)$, which satisfy a coupled system of first-order differential equations depending on the metric functions $f(R)$ and $g(R)$ \cite{FV2}.The Bogoliubov coefficients encoding the mixing phenomenon, denoted as $W_{p,j,\kappa}$, emerge from the curved-space inner products between modes associated with different mass eigenstates. These coefficients are structurally defined by the overlap integral of the radial functions corresponding to $M_1$ and $M_2$. Integrating these solutions into the energy-momentum tensor VEV yields the exact analytical expression for the energy density of the flavor vacuum:
\begin{eqnarray}\label{Eq.:ExplicitEnergyDensity}
	\nonumber \mathbb{T}_{00} &=&  4 \sin^2 \Theta \sqrt{f(R)} \sum_{L=1,2} \sum_{j,\kappa,m_j}\int_0^{\infty} dp \ E_L |W_{p,j,\kappa}|^2 \left(|\Phi_{L,p,j,\kappa}|^2 + |\Psi_{L,p,j,\kappa}|^2 \right)  \\
	&+& 4 \sin^2 \Theta \sqrt{f(R)} \sum_{L=1,2} \sum_{j,\kappa,m_j}\int_0^{M_L} dq \ E_L |W_{q,j,\kappa}|^2 \left(|\Phi_{L,q,j,\kappa}|^2 + |\Psi_{L,q,j,\kappa}|^2 \right)\ .
\end{eqnarray}
	Here, the summation runs over the mass eigenstates $L=1,2$ and the spherical quantum numbers $j$, $\kappa$, and $m_j$, which denote the total angular momentum, the spin-orbit coupling, and the z-component of the angular momentum, respectively. The integration variables $p = \sqrt{E_L^2 - M_L^2}$ and $q = \sqrt{M_L^2 - E_L^2}$ represent the continuous radial momenta for free states ($E_L \geq M_L$) and the pseudo-momenta for bound-like states ($E_L < M_L$). Noticeably, the energy density $\mathbb{T}_{00}$ is strictly positive and is directly proportional to $\sin^2 \Theta$, explicitly demonstrating that this dark matter-like contribution vanishes entirely in the absence of neutrino mixing.

\section{Galactic Dynamics from the Yukawa Potential}
To analyze the implications of the flavor vacuum on galactic scales, one considers static spherically symmetric spacetimes. In isotropic coordinates, the metric functions are given by $f(R)$ and $g(R)$. By applying the weak field approximation, the metric is expanded as $f(R) = 1 + 2 V(R)$ and $g(R) = 1 - 2V(R)$, where $V(R)$ is the gravitational potential treated as a small perturbation. Solving the Dirac equations in this background is highly non-trivial. However, approximate solutions can be constructed by perturbing the flat spacetime radial functions $\Phi^0_{L,p,j,\kappa}(R)$ and $\Psi^0_{L,p,j,\kappa}(R)$. To preserve the correct normalization of the modes up to linear order in $V(R)$ against the metric volume factor $g^{3/2}(R) \simeq 1 - 3V(R)$, the wavefunctions must be scaled by a compensating factor. The approximate radial functions are therefore given by:
\begin{equation}
	\Phi_{L,p,j,\kappa} (R) \simeq \left(1 + \frac{3}{2} V(R) \right) \Phi^0_{L,p,j,\kappa}(R) \ , \; \Psi_{L,p,j,\kappa} (R) \simeq \left(1 + \frac{3}{2} V(R) \right) \Psi^0_{L,p,j,\kappa}(R) \ .
\end{equation}
Inserting these modes into the explicit formula for the energy density \eqref{Eq.:ExplicitEnergyDensity}, and assuming the potential is sufficiently weak to prevent the formation of bound states ($E_L \leq M_L$), one can analytically evaluate the flavor vacuum contribution. The spatial metric determinant expansion and the wavefunction scaling lead to a total energy density $\mathbb{T}_{00}$ that is proportional to the flat spacetime condensation density $\mathcal{K}$, modified by a linear correction in the potential:
\begin{equation}
	\mathbb{T}_{00} \simeq 4 \sin^2 \Theta \ \mathcal{K} \left(1 + 4 V(R) \right) \ .
\end{equation}
Assuming the flavor vacuum is the primary energy-momentum source modifying the standard picture, the potential must satisfy the Poisson equation $\nabla^2 V = 4 \pi G \mathbb{T}_{00}$. Substituting the approximated density, the differential equation for the potential reads:
\begin{equation}
	V'' + \frac{2}{R}V' - \alpha (1 + 4V) = 0 \ ,
\end{equation}
where primes denote derivatives with respect to $R$, and we have defined the positive constant $\alpha = 16\pi G \mathcal{K} \sin^2 \Theta$. Discarding the unphysical diverging root for $R \rightarrow \infty$, the solution yields a Yukawa-type gravitational potential. For a galaxy with a total baryonic mass $\mathcal{M}$, the total effective gravitational potential acting on a test mass combines the standard Newtonian term generated by the visible matter with the flavor vacuum Yukawa correction. By appropriately setting the integration constant, the total potential is expressed as:
\begin{equation}\label{Eq.:TotalPotential}
	V(R) = -\frac{G\mathcal{M}}{R}\left( 1 + \beta e^{-\frac{R}{d}}\right) \ ,
\end{equation}
where $\beta$ is a dimensionless coupling parameter and $d = \frac{1}{2 \sqrt{\alpha}}$ is a characteristic scale length strictly related to the QFT ultraviolet cutoff $\Lambda$ embedded in $\mathcal{K}$. This modified potential directly affects the circular velocity $v(R)$ of visible matter. Equating the centripetal acceleration to the gravitational force, $v^2(R)/R = dV/dR$, one obtains:
\begin{equation}
	v^2 (R) = \frac{G\mathcal{M}}{R} \left[1 + \beta e^{-\frac{R}{d}}\left(1+ \frac{R}{d} \right) \right] \ .
\end{equation}
In the outer regions of the galaxy ($R \gg d$), the Newtonian term becomes negligible and the velocity profile flattens, approaching the constant asymptotic value:
\begin{equation}
	v^2_{\mathrm{FLAT}} \simeq \frac{G\mathcal{M}\beta}{d} \ .
\end{equation}
This behavior provides a mathematically consistent physical mechanism to account for the observed flat rotation curves without requiring ad hoc distributions of classical dark matter particles. Furthermore, the Yukawa potential framework naturally incorporates the empirical baryonic Tully-Fisher (BTF) relation:
\begin{equation}\label{BTF}
	v^4_{FLAT} = \xi \mathcal{M},
\end{equation} 
where $a_0 \simeq 1.2 \times 10^{-10} \mathrm{m/s^2}$ is a universal characteristic acceleration originally proposed in MOND phenomenology. Equating the theoretical flat velocity squared to the BTF requirement yields an explicit expression for the theoretical acceleration:
\begin{equation}
	a_{0} = \frac{4G\mathcal{M} \beta^2 e^{-2}}{d^2} \ .
\end{equation}
To ensure that $a_0$ remains a universal constant across different galaxies, specific scaling laws relating the parameters $\beta$ and $d$ (and consequently $\Lambda$) to the galactic baryonic mass $\mathcal{M}$ are required. Two viable phenomenological scenarios are analyzed and successfully tested against observational data from samples of spiral and gas-rich galaxies:
\begin{itemize}
	\item Universal $\beta$ and varying cutoff: The coupling $\beta$ is constant across galaxies, while the scale length $d$ varies. Since $d \propto \Lambda^{-1}$, this requires the QFT ultraviolet cutoff to scale with the galaxy mass as $\Lambda \sim \mathcal{M}^{-1/2}$.
	\item Universal cutoff and varying $\beta$: From a QFT perspective, a constant fundamental cutoff is more appealing. In this scenario, $\Lambda$ is strictly constant (e.g., set to the electroweak scale $\Lambda_{EW} \simeq 246 \ \mathrm{GeV}$), while the coupling parameter scales dynamically as $\beta \sim \mathcal{M}^{-1/2}$.
\end{itemize}

\begin{figure}[h]
	\begin{centering}
		\includegraphics[scale=0.43]{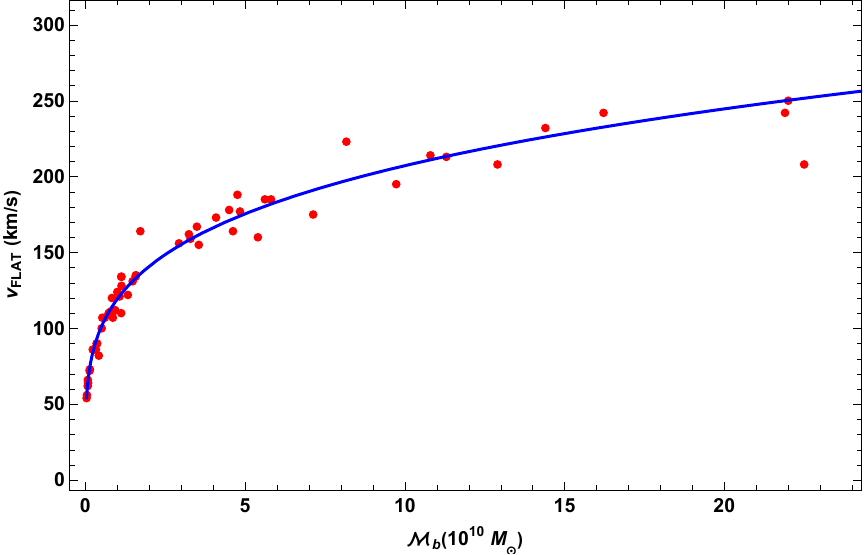} \hfill \includegraphics[scale=0.43]{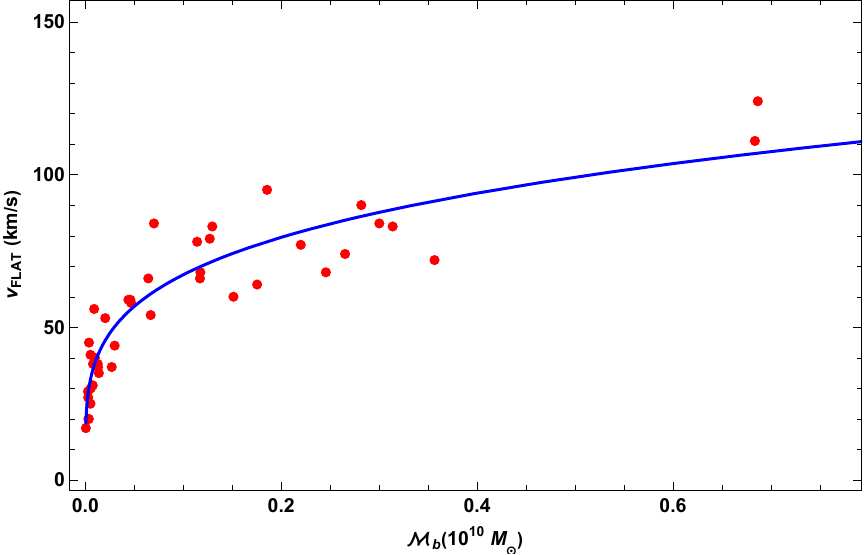}
		\includegraphics[scale=0.43]{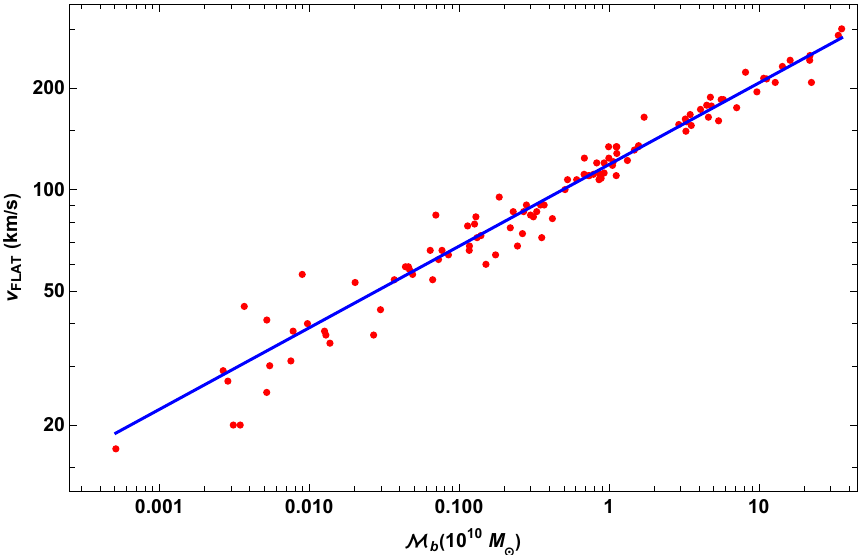}
		\par\end{centering}
	\caption{\label{Figura1} (color online) Plots of the best fit (solid blue line) for Eq. \eqref{BTF} for constant $\beta$ and cutoff scaling $\Lambda = \Lambda_0 \left(\frac{\mathcal{M}}{\mathcal{M}_{MW}} \right)^{\nu}$ versus the experimental points (red): (top left) spiral galaxies dataset (Data for spiral galaxies, extrapolated from \cite{DataSpirals}), (top right) gas rich galaxies dataset (Data for gas rich galaxies, extrapolated from \cite{DataGas}), (bottom) logarithmic scale plot for the combined dataset. The best fit parameters are respectively: spirals $(\beta = 0.433117, \nu = -0.520693)$, gas dominated $(\beta = 0.417015, \nu = -0.517175)$ and combined $(\beta = 0.426762, \nu = -0.515629)$. We have used the following values of neutrino masses $m_1 = 10^{-3} \ \mathrm{eV}$, $m_2 = \sqrt{m_1^2 + \Delta m_{12}^2}$, with $\Delta m_{12}^2 = 7.53 \times 10^{-5} \mathrm{eV}^2$ and the $e-\mu$ mixing angle $\sin^2 \theta_{12} = 0.307$ \cite{PDG24}.} 
\end{figure}
\begin{figure}[h]
	\begin{centering}
		\includegraphics[scale=0.43]{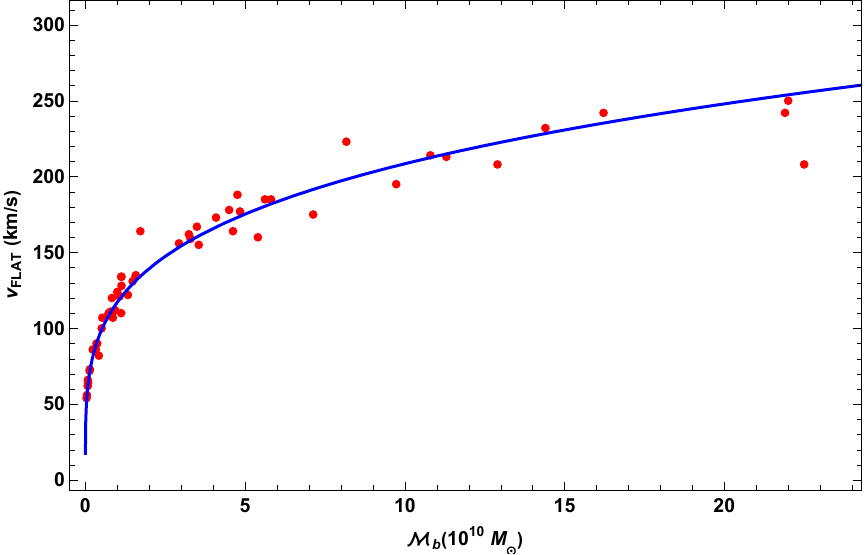} \hfill \includegraphics[scale=0.43]{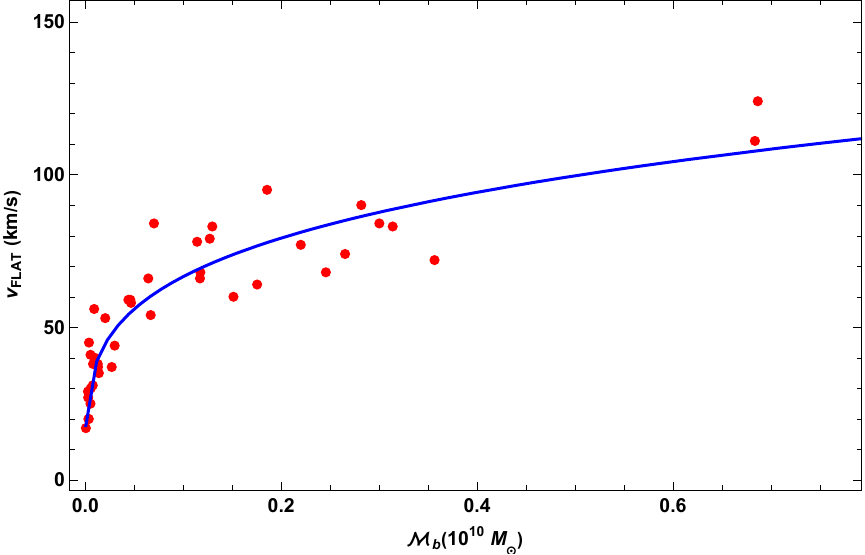}
		\includegraphics[scale=0.43]{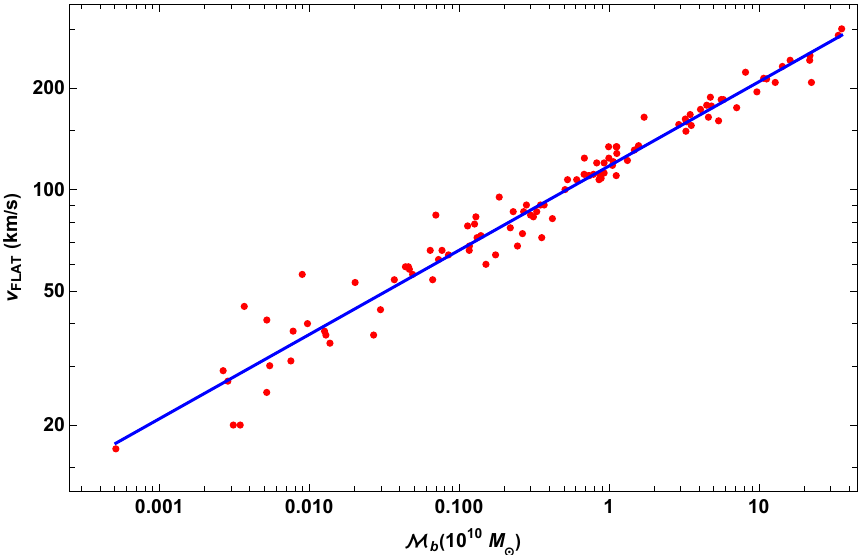}
		\par\end{centering}
	\caption{\label{Figura2} (color online) Plots of the best fit (solid blue line) for Eq. \eqref{BTF} for constant cutoff  $\Lambda = \Lambda_0 \simeq \ 15.425 \ \mathrm{keV}$ and $\beta$ scaling as $\beta = \beta_0 \left(\frac{\mathcal{M}}{\mathcal{M}_{MW}} \right)^{-\frac{1}{2}}$ versus the experimental points (red): (top left) spiral galaxies dataset (Data for spiral galaxies, extrapolated from \cite{DataSpirals}), (top right) gas rich galaxies dataset (Data for gas rich galaxies, extrapolated from \cite{DataGas}), (bottom) logarithmic scale plot for the combined dataset. The best fit parameters are respectively: spirals $\beta_0 = 0.413572$, gas dominated $\beta_0 = 0.422315$ and combined $\beta_0 = 0.414944$. The same parameters as Fig. 1 are used for neutrino masses and mixing angle.}
\end{figure}
\begin{figure}[h]
	\begin{centering}
		\includegraphics[scale=0.43]{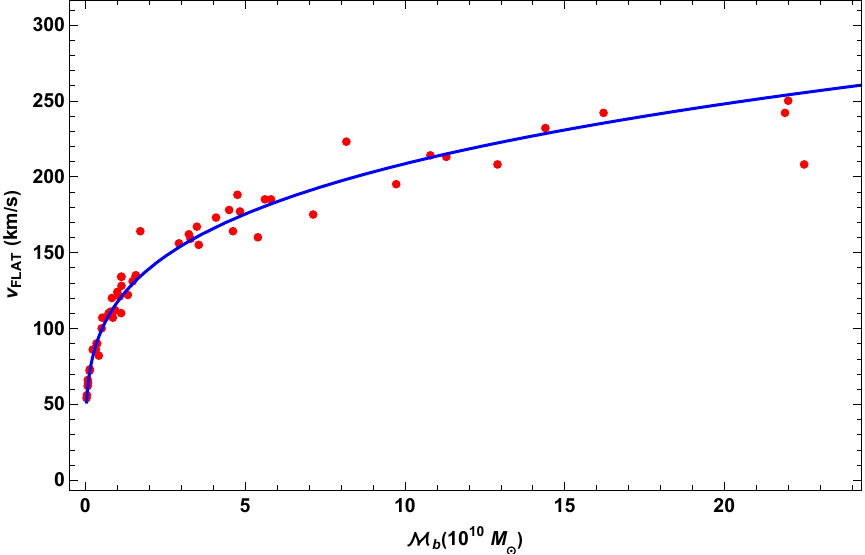} \hfill \includegraphics[scale=0.43]{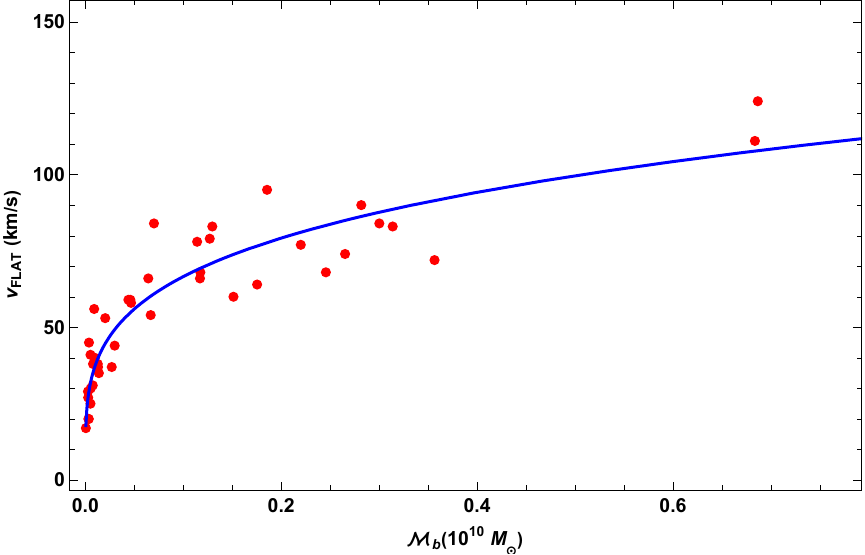}
		\includegraphics[scale=0.43]{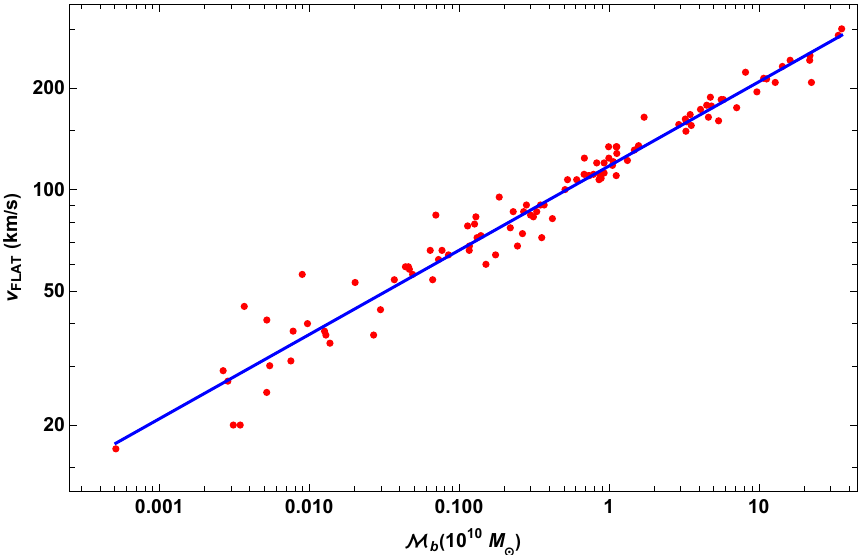}
		\par\end{centering}
	\caption{\label{Figura3} (color online) Plots of the best fit (solid blue line) for Eq. \eqref{BTF} for constant cutoff  $\Lambda = \Lambda_{EW} \simeq \ 246 \ \mathrm{GeV}$ and $\beta$ scaling as $\beta = \beta_0 \left(\frac{\mathcal{M}}{\mathcal{M}_{MW}} \right)^{-\frac{1}{2}}$ versus the experimental points (red): (top left) spiral galaxies dataset (Data for spiral galaxies, extrapolated from \cite{DataSpirals}), (top right) gas rich galaxies dataset (Data for gas rich galaxies, extrapolated from \cite{DataGas}), (bottom) logarithmic scale plot for the combined dataset. The best fit parameters are respectively: spirals $\beta_0 = 2.59436 \times 10^{-8}$, gas dominated $\beta_0 = 2.64805 \times 10^{-8}$ and combined $\beta_0 = 2.60183 \times 10^{-8}$.}
\end{figure}

\begin{equation}
	a_{0} = \frac{4G\mathcal{M} \beta^2 e^{-2}}{d^2} \ .
\end{equation}
To maintain a universal constant $a_0$, specific scaling laws for $\beta$ and the cutoff $\Lambda$ are required. Two viable phenomenological scenarios are successfully tested using data from samples of spiral and gas-rich galaxies:
\begin{itemize}
	\item Universal $\beta$ and varying cutoff: The QFT ultraviolet cutoff $\Lambda$ scales with the galaxy mass as $\Lambda \sim \mathcal{M}^{-1/2}$.
	\item Universal cutoff and varying $\beta$: The cutoff $\Lambda$ is strictly constant (e.g., set to the electroweak scale $\Lambda_{EW} \simeq 246 \ \mathrm{GeV}$), while the coupling parameter scales as $\beta \sim \mathcal{M}^{-1/2}$.
\end{itemize}
Both parameterizations yield excellent fits to the observational datasets, demonstrating that the neutrino flavor vacuum constitutes a flexible and rigorously derived theoretical framework for modeling galactic dynamics.

\section{Conclusions}
By employing quantum field theory in curved spacetime, we reviewed the general properties of the energy-momentum tensor associated with the neutrino flavor vacuum. It is established that this vacuum state attains a perfect fluid form characterized by an equation of state consistent with dust and cold dark matter ($p=0$). The explicit derivation in static, spherically symmetric spacetimes within the weak field limit shows that the flavor vacuum induces a Yukawa correction to the Newtonian gravitational potential. This theoretical result has direct phenomenological applications at galactic scales, where it naturally accounts for the flatness of spiral galaxy rotation curves. Moreover, by defining specific scaling relations for the theoretical parameters, the model accurately reproduces the baryonic Tully-Fisher relation. In this context, neutrino mixing acts as an intrinsic physical source that mimics MOND-like effects or extended gravity behaviors. These findings reinforce the interpretation of the flavor vacuum as a viable and theoretically grounded component of the dark matter sector.

\section*{Acknowledgments}
We acknowledge partial financial support from MUR and INFN. A.C. also acknowledges the COST Action CA1511 Cosmology and Astrophysics Network for Theoretical Advances and Training Actions (CANTATA) and COST Action CA21136 (CosmoVerse).

\section*{ORCID}

\noindent Antonio Capolupo - \url{https://orcid.org/0000-0002-8745-2522}

\noindent Salvatore Capozziello - \url{https://orcid.org/0000-0003-4886-2024 }

\noindent Gabriele Pisacane - \url{https://orcid.org/0009-0006-6626-6655}

\noindent Aniello Quaranta - \url{https://orcid.org/0000-0002-8190-4989}

\appendix

\end{document}